\documentclass[aps,prd,superscriptaddress,twocolumn,nofootinbib,amsmath,amssymb,nolongbibliography]{revtex4-2}

\usepackage{graphicx}
\usepackage{color}
\usepackage{url}
\usepackage{booktabs} 
\usepackage{lineno} 

\usepackage{dcolumn}

\usepackage{lipsum}
\usepackage{cuted}
\usepackage[normalem]{ulem} 

\usepackage{hyperref}
\definecolor{linkcolor}{rgb}{0,0,1}
\definecolor{urlcolor}{rgb}{0,0,.7}
\definecolor{citecolor}{rgb}{0,0,1}
\definecolor{acrocolor}{rgb}{0,0,.7}
\hypersetup{bookmarksopen,colorlinks=true}
\hypersetup{pdfstartview=FitH}
\hypersetup{linktocpage=true,bookmarksnumbered=true}
\hypersetup{plainpages=false,breaklinks=true}
\hypersetup{linkcolor=linkcolor,citecolor=citecolor,urlcolor=urlcolor}

\newcommand{\mr}[1]{\mathrm{#1}}
\definecolor{purple}{rgb}{0.855,0.439,0.839}

\begin{document}
	
\title{Characterisation of birefringence inhomogeneity of KAGRA sapphire mirrors from transmitted wavefront error measurements}

\author{Haoyu~Wang}
\email{haoyu.wang@phys.s.u-tokyo.ac.jp}
\affiliation{Research Center for the Early Universe (RESCEU), Graduate School of Science, University of Tokyo, Tokyo 113-0033, Japan}
\author{Yoichi~Aso}
\email{yoichi.aso@nao.ac.jp}
\affiliation{National Astronomical Observatory of Japan (NAOJ), Gravitational Wave Science Project, Tokyo 181-8588, Japan}
\author{Matteo~Leonardi}
\affiliation{Dipartimento di Fisica, Universit\`a di Trento, 38123 Povo, Trento, Italy}
\author{Marc~Eisenmann}
\affiliation{National Astronomical Observatory of Japan (NAOJ), Gravitational Wave Science Project, Tokyo 181-8588, Japan}
\author{Eiichi~Hirose}
\affiliation{Inhbar, Inc., Adachi, Tokyo 121-0051, Japan}
\author{GariLynn~Billingsley}
\affiliation{LIGO Laboratory, California Institute of Technology, Pasadena, California 91125, USA}
\author{Keiko~Kokeyama}
\affiliation{School of Physics and Astronomy, Cardiff University, Cardiff CF24 3AA, UK}
\author{Takafumi~Ushiba}
\affiliation{Institute for Cosmic Ray Research (ICRR), KAGRA Observatory, The University of Tokyo, Gifu 506-1205, Japan}
\author{Masahide Tamaki}
\affiliation{Institute for Cosmic Ray Research (ICRR), KAGRA Observatory, The University of Tokyo, Gifu 506-1205, Japan}
\affiliation{Department of Physics, University of Tokyo, Bunkyo, Tokyo 113-0033, Japan}
\author{Yuta~Michimura}
\affiliation{LIGO Laboratory, California Institute of Technology, Pasadena, California 91125, USA}
\affiliation{Research Center for the Early Universe (RESCEU), Graduate School of Science, University of Tokyo, Tokyo 113-0033, Japan}
\affiliation{PRESTO, Japan Science and Technology Agency (JST), Kawaguchi, Saitama 332-0012, Japan}

\date{\today}

\begin{abstract}
Cooling down test masses to cryogenic temperatures is a way to reduce the thermal noise of gravitational wave detectors.
Crystalline materials are considered the most promising materials for fabricating cryogenic test masses and their coatings because of their excellent thermal and optical properties at low temperatures. However, birefringence owing to local impurities and inhomogeneities in the crystal can degrade the performance of the detector. The birefringence measurement or mapping over a two-dimensional area is thus important.
This study describes a method for fast birefringence measurements of a large sample by simply combining a series of transmission wavefront error measurements using linearly polarised light with Fizeau interferometers. Using this method, the birefringence inhomogeneity information of KAGRA's two input test masses with a diameter of 22 cm was fully reconstructed. The birefringence information was then used to calculate the transverse beam shapes of the light fields in orthogonal polarisation directions when passing through the substrate. 
It was possible to find a calculated beam shape consistent with in-situ measurements using the KAGRA interferometer.
This technique is crucial for birefringence characterisation of test masses in future detectors, where even larger sizes are used.
\end{abstract}

\maketitle


\section{Introduction}

Thermal noise is one of the main noise sources for gravitational wave detectors~\cite{Flaminio2014, nawrodt2011challenges}. To reduce thermal noise, future detectors such as the Einstein Telescope~\cite{Punturo_2010, ET-0007B-20} and LIGO Voyager~\cite{Adhikari_2020} concept are designed to operate at cryogenic temperatures. Cosmic Explorer~\cite{hall2022cosmic, reitze2019cosmic} considers cryogenic operations for upgrades.
KAGRA~\cite{akutsu2021overview, michimura2020prospects, Somiya_2012, aso2013} is the only detector operating at cryogenic temperatures and pioneers this technique among the current large-scale ground-based detectors. All these detectors use large, high-quality mirrors as their test masses, forming optical resonators to sense the tiny length fluctuations introduced by gravitational waves. Fused silica is commonly used to make test masses at room temperature; however, it is not compatible with cryogenic detectors due to its large mechanical loss at low temperatures. Crystalline materials are considered to be the most promising materials for fabricating cryogenic test masses.
Sapphire was selected as the material for KAGRA's test masses~\cite{Hirose2014} because of its excellent thermal and optical properties at low temperatures~\cite{UCHIYAMA19995, Tomaru_2002}. 
However, recent studies have shown that the birefringence of crystalline substrates and coatings can degrade detector performance~\cite{Kruger2016, Hamedan2023, winkler2021, tanioka2023, somiya2019nonuniformity, michimura2023effects}.
The birefringence fluctuates spatially in magnitude and direction throughout the mirror substrate.
Therefore, it is necessary to measure the position dependence of the birefringence over a large area to assess its impact on optical performance.

There are several studies on two-dimensional birefringence measurements of transparent materials~\cite{Kruger2016, Hamedan2023, Pastrnak1971, Shribak2003, onuma2014, Villele2000, chu2002, Benabid1998}. These measurements use either a single-beam scanning technique or expand the beam to a large aperture for imaging using a sensor array. The scanning method is rather time-consuming for large sample sizes, for example, mirrors with diameters usually greater than 20~cm for ground-based gravitational wave detectors, if one wants to obtain good spatial resolution~\cite{Tokunari2006, zeidler2022simultaneous}. A phase-shifting method in digital interferometry has been proposed for fast 2D birefringence measurements with high accuracy~\cite{Noguchi1992, Otani1994, Cochran1992}. Fizeau interferometry is among the most accurate methods for extracting the phase information of an optical component (reflection or transmission phase) over a large area. The Transmission Wavefront Error (TWE) is often used to measure the refractive index homogeneity of a substrate under test. However, in the presence of birefringence a simple TWE measurement does not reveal all the information about the substrate.

In this study, we describe a method for combining a series of TWE map measurements using linearly polarised light in an appropriate way to fully reconstruct refractive index information including birefringence, which is introduced in Section~\ref{sec: Generating a birefringence map with a Fizeau interferometer}.
Using this method, the birefringence inhomogeneity information of KAGRA's two input test masses (ITM) is extracted, described in Section~\ref{sec: Birefringence characterisation of KAGRA sapphire ITMs}. This birefringence information was then used to calculate the transverse beam shape of the light fields in orthogonal polarisation directions when passing through the substrate, which is compared with the in-situ beam shape measurements.
Conclusions and discussions are summarised in Section~\ref{sec: Conclusion}.

\section{Generating a birefringence map with a Fizeau interferometer}\label{sec: Generating a birefringence map with a Fizeau interferometer}

\begin{figure}[htbp]
\includegraphics[width=8.5cm]{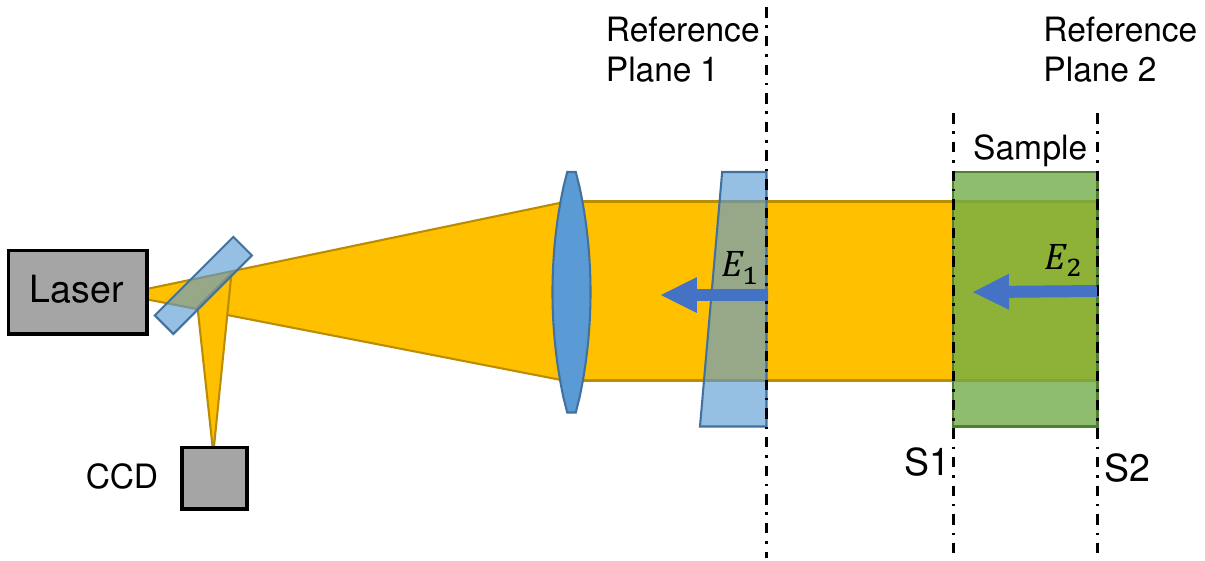}
\caption{Schematic of TWE measurement with a Fizeau interferometer. S1 is the surface with anti-reflection coating and S2 is the surface with highly reflective coating. The interference between light fields $E_\mathrm{1}$ and $E_\mathrm{2}$ reflected by reference planes 1 and 2, respectively, obtains the TWE information of the sample.}
\label{Fig: Fizeau Interferometer}
\end{figure}

Figure\,\ref{Fig: Fizeau Interferometer} shows a schematic diagram of the Fizeau interferometer. We analyse the interference pattern between the light reflected by the reference plane 1\,($E_\mathrm{1}$) and the light reflected by the reference plane 2\,($E_\mathrm{2}$) on the CCD. By comparing the interference between reference planes 1 and 2, we obtained information on the optical thickness of the sample. Because the optical thickness is the physical thickness multiplied by the index of refraction, by measuring the surface figure maps of S1 (surface with anti-reflection coating) and S2 (surface with highly reflective coating) precisely and subtracting the variation in the physical thickness of the sample, we can extract the fluctuation of the index of refraction in the substrate.

To extract birefringence information, measurements must be performed using linearly polarised light. To understand the behaviour of such measurements, we first considered the behaviour of linearly polarised light passing through a birefringent substrate.

\hspace*{\fill} \

\hspace*{\fill} \

\subsection{Transmission of a linearly polarised light through a birefringent medium}

\begin{figure}[htbp]
\includegraphics[width=6cm]{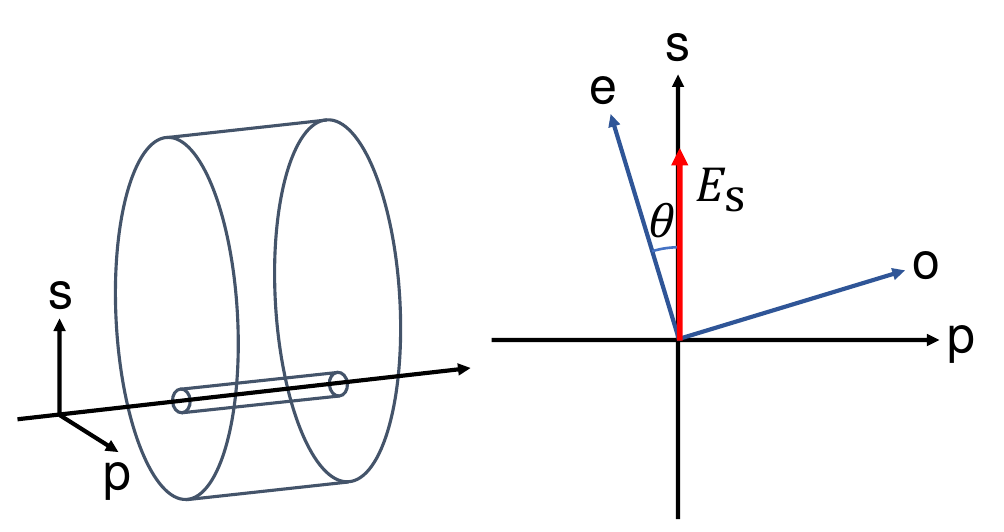}
\caption{Transmission of a linearly polarised optical field through a column of a sample substrate.}
\label{Fig: Jones Matrix}
\end{figure}

Consider optical axis passing through a single substrate column, as shown in Figure~\ref{Fig: Jones Matrix}. We defined our preferred polarisation axes, s and p, in the vertical and horizontal directions with respect to the laboratory frame. The extraordinary and ordinary axes of the crystal in that particular column are assumed to be rotated by an angle $\theta$ with respect to our polarisation axes. The input beam to the Fizeau interferometer was purely polarised in s ($E_\mr{s}$).

The Jones matrix formalism was used to analyse the conversion of the polarisation state\,\cite{Yariv}. The polarisation state of light is represented by the following vector:
\begin{equation}
\vec{V}=
    \begin{pmatrix}
        E_\mr{s}\\
        E_\mr{p}
    \end{pmatrix},
\end{equation}
where $E_\mr{s}$ and $E_\mr{p}$ are the complex amplitudes of the s- and p-polarised electric fields, respectively. After passing through the substrate twice (a round trip through the substrate), the polarisation state was converted to:
\begin{widetext}
\begin{equation}
\vec{V}' = M\cdot \vec{V},
\quad\mathrm{with}\quad
M = e^{i\alpha_+}
\begin{pmatrix}
    e^{i\alpha_-}\cos^2\theta + e^{-i\alpha_-}\sin^2\theta & -i\sin\alpha_-  \sin2\theta\\
    -i\sin\alpha_- \sin2\theta & e^{-i\alpha_-}\cos^2\theta + e^{i\alpha_-}\sin^2\theta
\end{pmatrix},\label{eq: Jones matrix}
\end{equation}
\end{widetext}
where $M$ denotes the round-trip Jones matrix of the substrate. The indices of refraction for the ordinary and extraordinary axes are denoted as $n_\mr{o}$ and $n'_\mr{e}$~\cite{Tokunari2006}, respectively. Then, we can compute the round- trip phase changes of the optical fields polarised along the axes:
\begin{equation}
    \alpha_\mr{o} \equiv 2\pi\frac{2dn_\mr{o}}{\lambda},
\end{equation}
\begin{equation}
    \alpha_\mr{e} \equiv 2\pi\frac{2dn'_\mr{e}}{\lambda},
\end{equation}
where $d$ is the thickness of the substrate, and $\lambda$ is the wavelength of the light.

Using $\alpha_\mr{o}$ and $\alpha_\mr{e}$, we define the common and differential phase changes as follows:
\begin{equation}
    \alpha_+ \equiv \frac{\alpha_\mr{o} + \alpha_\mr{e}}{2} = \frac{4\pi d}{\lambda} (n_\mr{o} + n'_\mr{e}),
\end{equation}
\begin{equation}
    \alpha_- \equiv \frac{\alpha_\mr{o} - \alpha_\mr{e}}{2} = \frac{4\pi d}{\lambda} (n_\mr{o} - n'_\mr{e}).
\end{equation}
In most cases, the common phase is not important and we will ignore $\alpha_+$ in the following discussion. Note that the Jones matrix also works for a non-uniform medium (discussed in Appendix~\ref{app: Effective Jones matrix of a non-uniform medium along the optical axis}).

For a purely s-polarised input field, $\vec{V} = (1, 0)^\mr{T}$, the field returning from the substrate becomes,
\begin{equation}
    \vec{V}' =
    \begin{pmatrix}
        E_\mr{s}\\
        E_\mr{p}
    \end{pmatrix} =
    \begin{pmatrix}
        e^{i\alpha_-}\cos^2\theta + e^{-i\alpha_-}\sin^2\theta\\
        -i\sin\alpha_- \sin2\theta
    \end{pmatrix}.
    \label{eq: V'}
\end{equation} 
This $\vec{V}'$ is the same as the field $E_2$, as shown in Figure~\ref{Fig: Fizeau Interferometer}, at the CCD, except for the common phase term, which is ignored.

\subsection{Interference with the reference field}
The field $\vec{V}'$ computed above is superposed with the reference field ($E_1$ in Figure~\ref{Fig: Fizeau Interferometer}). The reference field $\vec{V}_\mr{r}$ is purely polarised in s and has a certain phase offset $\phi$ from $\vec{V}'$,
\begin{equation}
    \vec{V}_\mr{r} =
    \begin{pmatrix}
        E_\mr{r}\\
        0
    \end{pmatrix} =
    \begin{pmatrix}
        e^{i\phi}\\
        0
    \end{pmatrix}.
    \label{eq: Vr}
\end{equation}

The optical power detected by the CCD after the interference of $\vec{V}'$ and $\vec{V}_\mr{r}$ is as follows:
\begin{equation}
    P = (\vec{V}_\mr{r} + \vec{V}')\cdot (\vec{V}_\mr{r} + \vec{V}')^*.
    \label{eq: P}
\end{equation}

By substituting Equations~(\ref{eq: V'}) and (\ref{eq: Vr}) into Equation~(\ref{eq: P}), we obtain:
\begin{align}\notag
    P & = |E_\mr{s} + E_\mr{r}|^2 + |E_\mr{p}|^2\\
    & = |E_\mr{s}|^2 + |E_\mr{p}|^2 + |E_\mr{r}|^2  + E_\mr{s}E_\mr{r}^* +  E_\mr{s}^*E_\mr{r}.
\end{align}
Based on energy conservation, $|E_\mr{s}|^2 + |E_\mr{p}|^2 = 1$. From Equation~(\ref{eq: Vr}), $|E_\mr{r}|^2=1$. We ignore these constant terms in the following discussion. However, note that
\begin{align}\notag
    \Delta P &=  E_\mr{s}E_\mr{r}^* +  E_\mr{s}^*E_\mr{r}\\
    &= 2\cos^2\theta\cos(\phi - \alpha_-) + 2\sin^2\theta\cos(\phi + \alpha_-).
\end{align}
If the s-polarisation axis is aligned with the extra-ordinary axis ($\theta=0$), $\Delta P = 2\cos(\phi - \alpha_-)$. A Fizeau interferometer is equipped with a mechanism to change the position of the reference plane 1, for example, by mounting a reference flat on a piezo stage. Therefore, we can scan the relative phase $\phi$ by known amounts and fit the measured variation of $\Delta P$ to obtain the value of $\alpha_-$. In the general case of $\theta\neq 0$, the Fizeau measurement yields the phase of the field $E_\mr{s}$, that is, $\psi = \arg(E_\mr{s})$.

\subsection{Extraction of $\alpha_-$ and $\theta$ from several TWE maps}\label{sec: Extraction}

When the orientation of the extraordinary axis is unknown, which is almost always the case, we must obtain several TWE maps with different orientations of the sample or measurement beam polarisation. Then combine them to extract $\theta$ and $\alpha_-$.

As explained in the previous section, a TWE measurement by a Fizeau interferometer yields a transmission phase $\psi$ at each point on the sample surface. Let us assume that we rotate the sample by an angle $\eta$, or equivalently the input polarisation orientation by $-\eta$. This is equivalent to changing the orientation of the extra-ordinary axis from $\theta$ to $\theta + \eta$.

Let us denote the measured phase with an $\eta$ rotation of the sample as $\psi(\eta)$. From Equation~(\ref{eq: V'}), 
\begin{align}\notag
    \psi(\eta) &= \arg\left[E_\mr{s}(\theta+\eta)\right]\\\notag
    &= \arg\left[\cos^2(\theta+\eta)e^{i\alpha_-} + 
    \sin^2(\theta+\eta)e^{-i\alpha_-}\right]\\
    &=\arctan\left[\cos(2\theta+2\eta)\tan\alpha_-\right].\label{eq: psi eta}
\end{align}
We can see $\psi(\eta)$ is a function of $\eta$ with parameters $\theta$ and $\alpha_-$ determining its shape. Therefore, if we have several measurements of $\psi(\eta)$ with different values of $\eta$, we can use a non-linear fitting algorithm to fit the measured values with Equation~(\ref{eq: psi eta}) to obtain the best fit values of $\theta$ and $\alpha_-$.

If $\alpha_-$ can be assumed to be very small, which is the case for KAGRA mirrors, we can expand Equation~(\ref{eq: psi eta}) with $\alpha_-$, leaving only the $\mathcal{O}(\alpha_-)$ terms, as follows:
\begin{align}\notag
    \psi(\eta) &\sim \cos(2\theta+2\eta)\alpha_-\\
    &=\left(\cos(2\theta)\cos(2\eta) - \sin(2\theta)\sin(2\eta)\right)\alpha_-. 
\end{align}

Assuming that we have four measurements of $\psi(\eta)$ with four different values of $\eta = \eta_1, \eta_2, \eta_3,\eta_4$, we compute the following quantities,
\begin{align}
    \Delta\psi_1 &\equiv \psi(\eta_1) - \psi(\eta_2) =\alpha_-\left(A\cos2\theta - B\sin2\theta\right)\\
    \Delta\psi_2 &\equiv \psi(\eta_3) - \psi(\eta_4) =\alpha_-\left(C\cos2\theta - D\sin2\theta\right), 
\end{align}
where,
\begin{align}
    A &= \cos2\eta_1 - \cos2\eta_2\\
    B &= \sin2\eta_1 - \sin2\eta_2\\
    C &= \cos2\eta_3 - \cos2\eta_4\\
    D &= \sin2\eta_3 - \sin2\eta_4. 
\end{align}
We then use the ratio of $\Delta\psi_1$ and $\Delta\psi_2$,
\begin{align}\label{eq: r}
    r &\equiv \frac{\Delta\psi_1}{\Delta\psi_2} = \frac{A\cos2\theta - B\sin2\theta}{C\cos2\theta - D\sin2\theta}=\frac{A - B\tan2\theta}{C - D\tan2\theta}.
\end{align}
Solving the above for $\theta$,
\begin{equation}\label{eq: theta}
    \theta = \frac{1}{2}\arctan\left(\frac{A-rC}{B-rD}\right).
\end{equation}
Once the value of $\theta$ is obtained, we can use Equation~(\ref{eq: psi eta}) to compute the value of $\alpha_-$ as follows:
\begin{equation}
    \alpha_- = \arctan\left[\frac{\tan\left\{\psi(\eta)\right\}}{\cos(2\theta+2\eta)}\right].
\end{equation}

\section{Birefringence characterisation of KAGRA sapphire Input test masses}\label{sec: Birefringence characterisation of KAGRA sapphire ITMs}

\begin{table}[b]
\caption{\label{tab: TWE specification}
Rms values of TWE specifications for the KAGRA test masses and measurements with a circular/linear polarisation beam within an aperture of 140~mm~\cite{Hirose2020}.
}
\begin{ruledtabular}
\begin{tabular}{cccc}
 &  & Circular & Linear \\
 & Specification &  polarisation &  polarisation \\
\hline
ITMX & $<$5~nm & 4.09~nm & 25.90~nm\\
ITMY & $<$5~nm & 4.61~nm & 32.72~nm\\
\end{tabular}
\end{ruledtabular}
\end{table}

Sapphire is a uniaxial crystal with an optical axis called the c-axis that has no intrinsic birefringence. KAGRA’s sapphire test masses are fabricated such that the input surface is perpendicular to the c-axis. Light propagating in the test mass substrate travels along the c-axis, and theoretically, should experience no birefringence. However, local impurities and inhomogeneities in the crystal, as well as mechanical and thermal stress (e.g. due to the suspension system supporting the mirror), can cause birefringence. These local defects distorted the polarisation distribution of the light field in the transverse plane.

Gravitational wave detectors have very strict requirements on the TWE of light going through input test mass substrates. To achieve this requirement and minimise the influence of crystal inhomogeneity, an ion beam figuring technique was used during the polishing process, in which the thickness of the substrate was precisely adjusted over the surface to ensure that the TWE rms was within a few nanometres~\cite{somiya2019nonuniformity}. However, this technique optimizes the TWE for the beam with a certain polarisation state, and it cannot reduce birefringence.

\subsection{TWE of KAGRA sapphire test mass}

\begin{figure}[htbp]
	\begin{center}
		\includegraphics[width=8.5cm]{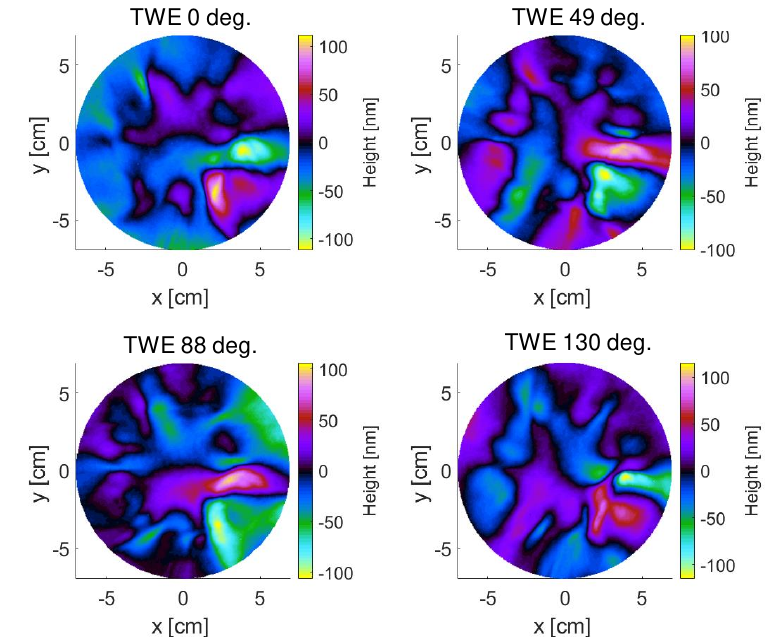}
		\caption{TWE maps of KAGRA's current ITMY within the 140 mm aperture measured with linearly polarised light with different $\theta$.}
		\label{fig: TWE ITMY}
	\end{center}
\end{figure}

The sapphire ITMs have already been installed in KAGRA, and direct birefringence mapping is impossible now. However, characterisation of the sapphire test masses was studied in~\cite{Hirose2020}. The TWE maps of the ITMs were measured several times with linearly polarised light by rotating the sample at different angles, which was equivalent to the measurements obtained by rotating the input polarisation. The TWE maps were measured at the angles of 0$^{\circ}$, 45$^{\circ}$, 90$^{\circ}$ and 135$^{\circ}$. The methods described in Section~\ref{sec: Generating a birefringence map with a Fizeau interferometer} require a rigid angular control. 
However, the main motivation of the measurements in~\cite{Hirose2020} was to see the stress induced change in $\alpha_+$, not $\alpha_-$, by different mounting orientations. Therefore,  the rotation angle was not precisely controlled.
To mitigate the problem, we compared the locations of the markers on the edge of the mirrors in different maps. In this way, we were able to recover the real rotation angle (Figure~\ref{fig: TWE ITMY}). The TWEs of the KAGRA current ITMs measured with linearly polarised light are around 5 to 6 times larger than the specifications because the vendor used circularly polarised light to measure the path length difference and calculate the removal from surface S2 during the ion beam figuring process (rms values are listed in Table~\ref{tab: TWE specification}). As shown in Figure~\ref{fig: TWE ITMY}, the 0$^{\circ}$ map and the 88$^{\circ}$ map have almost opposite values (the same is true for 49$^{\circ}$ and 130$^{\circ}$ maps). By summing the two maps, we can derive the TWE for circularly polarised light, whose rms magnitude is below 5~nm and matches with the specification.

\subsection{Constructing birefringence maps from TWE}\label{sec: Constructing birefringence maps from TWE}

The biggest challenge for the birefringence characterisation of KAGRA’s current ITMs is that the previous TWE measurement setup was not optimised for birefringence studies. According to Equation~(\ref{eq: r}) in Section~\ref{sec: Extraction}, to extract the birefringence information, we need to subtract two pairs of TWE maps and then divide one by the other. In typical Fizeau interferometer measurements, low-order effects such as the piston and surface tilt are removed from the results.
When doing measurements with different $\theta$, the distance and alignment between the sample and reference mirror have changed after rotating the sample. This implies that the difference between the two maps contains an unknown piston and tilt. The numerator and denominator in Equation~(\ref{eq: r}), which contain these unknown terms, provide incorrect birefringence information.

\begin{figure}[htbp]
\begin{center}
\includegraphics[width=8.5cm]{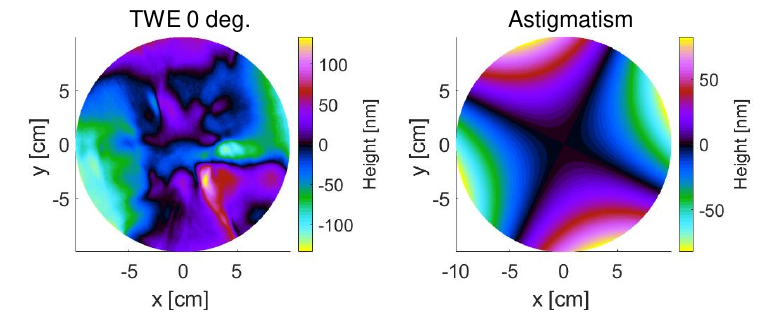}
\caption{Raw TWE map containing astigmatism which mainly occurs due to deformation due to gravity. As it is not in the vertical direction, we assume that the surface of the reference sphere used in the Fizeau interferometer is slightly ellipsoidal. After removing the astigmatism, the TWE is shown in the top left plot in Figure~\ref{fig: TWE ITMY}. The algorithm used in processing the maps was based on Simtools~\cite{simtools}.}
\label{fig: rawTWE}
\end{center}
\end{figure}

The centre of the map also changed after the sample was rotated. However, we observed that the decentring was very small (less than 2~mm), and its effect was negligible compared with the effects of the piston and tilt. In addition, astigmatism was observed on the original TWE maps. The left plot in Figure~\ref{fig: rawTWE} shows the TWE map after removing the curvature. An astigmatism term can be observed on the map plotted on the right. We observed that the rotation of this astigmatism did not change when the samples were rotated. Therefore, we believe that the astigmatism originates from the gravity-induced deformation of the sample and the astigmatism of the reference mirror used for the Fizeau interferometer. Astigmatism was individually removed from each TWE map.

\begin{figure}[htbp]
\begin{center}
\includegraphics[width=8.5cm]{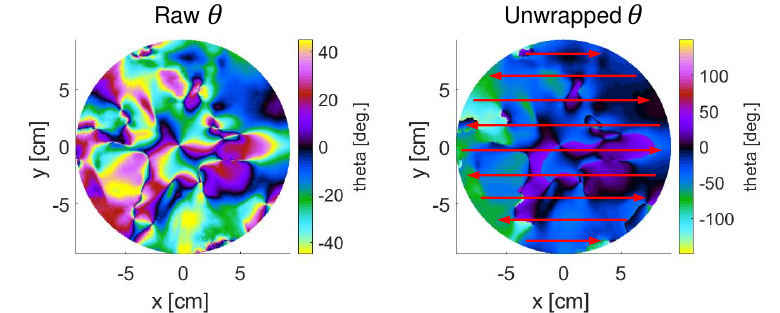}
\caption{The left plot shows initially derived $\theta$ map according to Equation~(\ref{eq: r}). The unknown piston and tilt will give wrong $\theta$ and cause failure of unwrapping shown on the right where the unwrapped surface is discontinuous. When performing parameter search for pistons, the 2D matrix is expanded to 1D vector along the arrow direction. Each arrow represents a row in the 477$\times$477 data matrix.}
\label{fig: unwrapping}
\end{center}
\end{figure}

From Equation~(\ref{eq: theta}), $\theta$ is wrapped in the range of ($-45^{\circ}$, $45^{\circ}$) due to the $\arctan$ function and we need to unwrap the data. Phase unwrapping techniques are widely used in optical metrology and interferometry~\cite{Judge1994, Su2004}. In our case, the problem of the unknown piston and tilt is that it will give the wrong $\theta$ and cause failure of unwrapping. Some algorithms improve phase unwrapping robustness against noise~\cite{Cusack95, Buckland95, Volkov03, Estrada11}. Most of these methods require the noise to be small compared with the phase change. However, the unmeasured piston and tilt on the TWE map can be large. As we take the division of two subtractions, the error will be further amplified when the denominator in Equation~(\ref{eq: r}) is close to zero. The left plot in Figure~\ref{fig: unwrapping} shows our initially derived $\theta$. The unwrapping algorithm does not function properly for such maps (shown on the right). This resulted in a discontinuous, unwrapped surface. The algorithm used in this study was based on~\cite{Herraez2002, kasim2017matlab}.

\begin{figure}[htbp]
	\begin{center}
		\includegraphics[width=8.5cm]{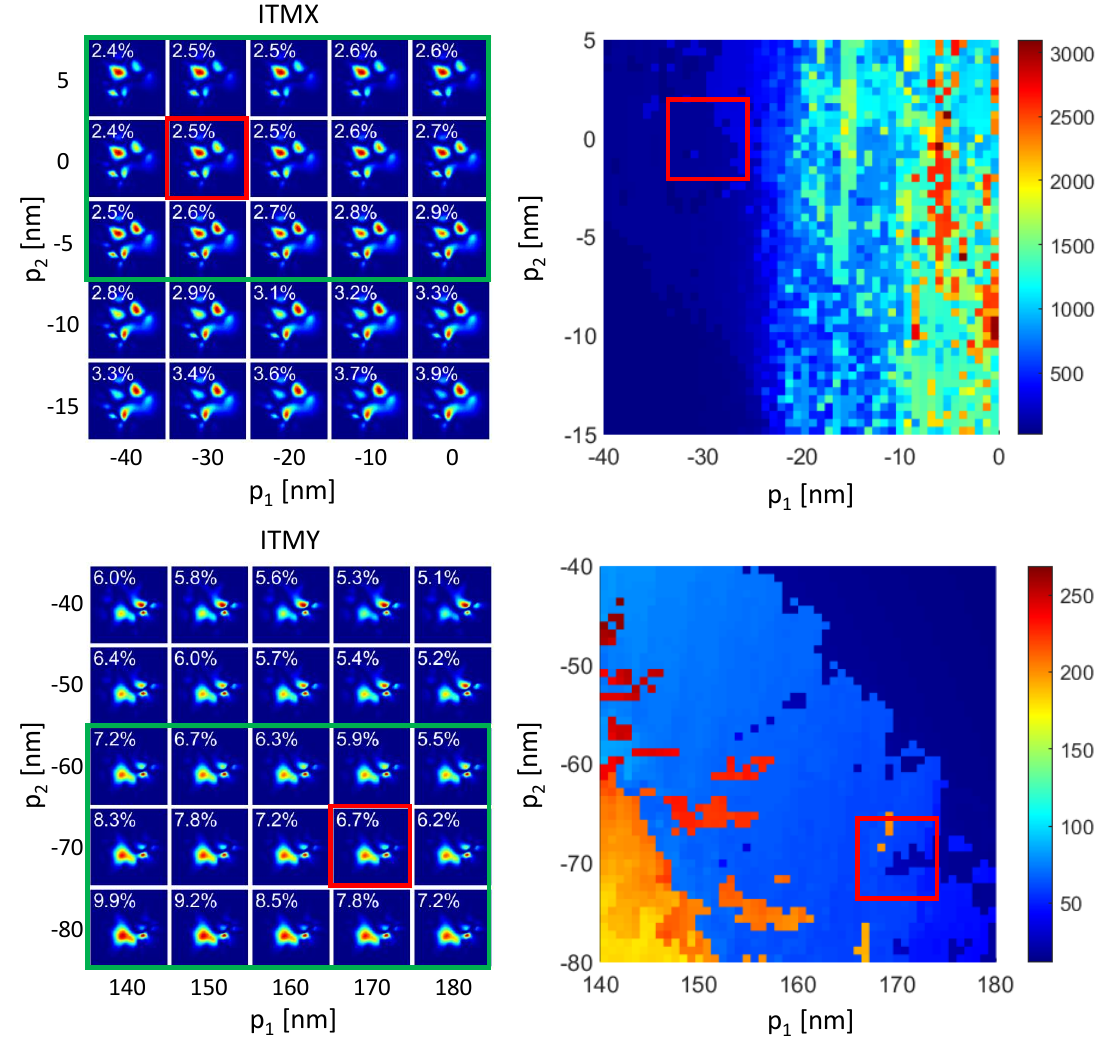}
		\caption{Parameter scan for piston term for ITMX (top) and ITMY (bottom). Plots on the left are single-bounce reflected (passing through the mirror substrate twice) beam shapes in p-polarisation (the input is pure s-polarisation) with different pistons applied to subtractions of TWEs. Numbers in white are the ratio percentage of beam power in p-polarisation over the total input power. Plots on the right are continuity factors of $\theta$ map after unwrapping. It can be observed that $p_1$ mainly influences the continuity factor, whereas $p_2$ determines the beam shape and power. We search for the correct pistons by eye where continuity factors are small and beam shapes match with measurements shown in Figure~\ref{fig: beamshape}. We assume that the beam shape shown in the red square gives the best estimation as a consideration of the continuity factor and measured beam shape. However, beams in the green area all have similar shapes with significantly varying power.}
		\label{fig: piston}
	\end{center}
\end{figure}

To recover the correct $\theta$, we perform parameter search for piston terms in the numerator and denominator shown in Equation~(\ref{eq: r}) Tilt terms are removed individually for each TWE map (by removing first-order Zernike polynomials) and ignored here. This yields
\begin{equation}\label{eq: new r}
	r = \frac{\psi(\eta_1) - \psi(\eta_2)+p_1}{\psi(\eta_3) - \psi(\eta_4)+p_2},
\end{equation}
where $p_1$ and $p_2$ are missing piston terms. For each pair of $p_1$ and $p_2$ values, the derived raw $\theta$ map was unwrapped. Here, we assume that unmeasured pistons are the main cause of unwrapping failure. The two-dimensional unwrapped map is then expanded to one-dimensional data along the arrow direction, as shown on the right in Figure~\ref{fig: unwrapping}. If the unwrapped map is a continuous surface, the one-dimensional data should be a continuous curve. Therefore, the overall continuity factor of the map can be simply described by
\begin{equation}\label{eq: smoothness}
	\mathcal{C} = \sum_{k=1}^{n}(\epsilon_{k+1}-\epsilon_{k})^2,
\end{equation}
where $\epsilon_{k}$ is the TWE value at each point of the one-dimensional data and $n$ is the total number of the data. A lower value of $\mathcal{C}$ indicates a smoother surface, whereas a higher value of $\mathcal{C}$ means that more values jump over the surface (refer the right plot shown in Figure~\ref{fig: unwrapping}). The continuity factors of ITMX and ITMY are shown on the right side of Figure~\ref{fig: piston}. For each pair of $p_1$ and $p_2$, $\theta$ and $\alpha_-$ can be derived. Subsequently, we calculate the cross-coupling term $E_\mr{p}=-i\sin\alpha_- \sin2\theta$ in Equation~(\ref{eq: Jones matrix}). This provided a field of linearly polarised light in the direction perpendicular to the input polarisation. The beam shape is shown on the left in Figure~\ref{fig: piston} with different piston values. 

\subsection{Comparison with in-situ measurements}

\begin{figure}[htbp]
\begin{center}
\includegraphics[width=8.5cm]{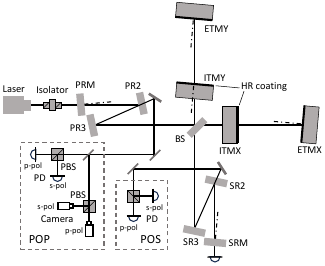}
\caption{Simplified schematic of KAGRA configuration.}
\label{fig: setup}
\end{center}
\end{figure}

Birefringence measurements were performed for the sapphire ITMs at KAGRA. A simplified schematic of the KAGRA configuration used during this measurement is shown in Figure~\ref{fig: setup}. The laser after the Faraday Isolator was in pure s-polarisation. The power recycling mirror (PRM), signal recycling mirror (SRM), and end test masses (ETMX and ETMY) are all misaligned such that the beam is reflected only by the ITMs. The figure shows the particular setup for measuring the birefringence of ITMX, where ITMY is misaligned. The coatings of the beam splitter (BS) were optimised for s-polarisation with a 50\%:50\% ratio for reflection and transmission at 45$^\circ$ incidence. The p-polarisation ratio was 20\%:80\% according to the coating design. Because there is no direct measurement of coating reflectivity for p-polarisation, the practical reflection and transmission may be different from the design values. The beams reflected from the ITM were measured at the POP and POS ports. Photodetectors (PD) and cameras are used for both s-polarisation and p-polarisation at the POP, whereas only photodetectors are  used at the POS. 

\begin{table}[b]
\caption{\label{tab:p-pol power}
Comparison between the measured beam power in p-polarisation at POP and POS ports with calculations from $\theta$ and $\alpha_-$ maps. Numbers are percentages of the p-polarisation power with respect to the total power reflected from the ITMX or ITMY.
}
\begin{ruledtabular}
\begin{tabular}{cccc}
 & POP & POS & Calculation \\
\hline
ITMX & $(3.65\pm0.02)\%$ & $(4.98\pm0.01)\%$ & 2.5\%\\
ITMY & $(8.07\pm0.04)\%$ & $(11.13\pm0.02)\%$ & 6.7\%\\
\end{tabular}
\end{ruledtabular}
\end{table}

The measured beam power in p-polarisation (as a percentage of the total reflected power) is listed in Table~\ref{tab:p-pol power}, where the effect of the unbalanced reflection and transmission of the BS on the p-polarisation is calibrated (according to the coating design values).
In principle, the measured beam power content in p-polarisation should be the same for the POP and POS ports. However, our measurements are different from each other. One reason for this could be that the practical reflection and transmission of the BS for p-polarisation deviate from the design. Another reason could be clipping and ghost beams. When the measurement was performed, gate valves with diameters slightly larger than the beam diameter were added to the vacuum tubes. We also found many ghost and scattering beams at the POP and POS. Although we dumped most of them, there may be still some remaining and are seen by the photodetectors.

\begin{figure}[htbp]
\begin{center}
\includegraphics[width=7cm]{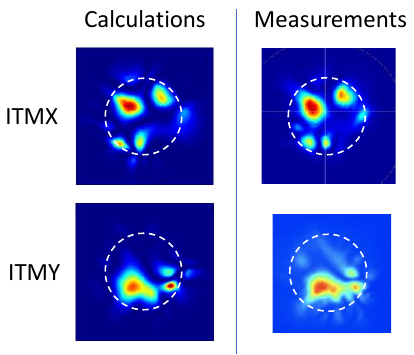}
\caption{Comparison between calculated p-polarisation beam shape and KAGRA measurements. Dashed circles in white represent the Gaussian beam diameter, the width at which the amplitude is $1/e$ of its value in the centre.}
\label{fig: beamshape}
\end{center}
\end{figure}

The measured beam shapes are shown on the right side in Figure~\ref{fig: beamshape}.
By comparing our calculated beam shapes from the previous section with practical measurements, we can find the correct pair of $p_1$ and $p_2$ visually, which gives both a low continuity factor and good agreement with the measured beam shapes (refer Figure~\ref{fig: piston}).
The beam power in p-polarisation was then calculated (the values are listed in Table~\ref{tab:p-pol power}) using the most suitable pair of $p_1$ and $p_2$ (marked with the red square shown in Figure~\ref{fig: piston}). The beam power estimated from our maps was lower than that from the measurements. One reason is that the unknown tilt term in the measured TWE maps influences the beam power. By adding tilt influences in the parameter scan, it is possible to increase the calculated power in p-polarisation while maintaining the beam shape unchanged (or slightly changed). In this study, we did not  perform a full parameter scan for tilt because the purpose was to demonstrate the feasibility of the method of extracting birefringence information from TWE measurements. An example of the influence of the tilt is presented in Appendix~\ref{app: tilt}.

\begin{figure}[htbp]
\begin{center}
\includegraphics[width=7.5cm]{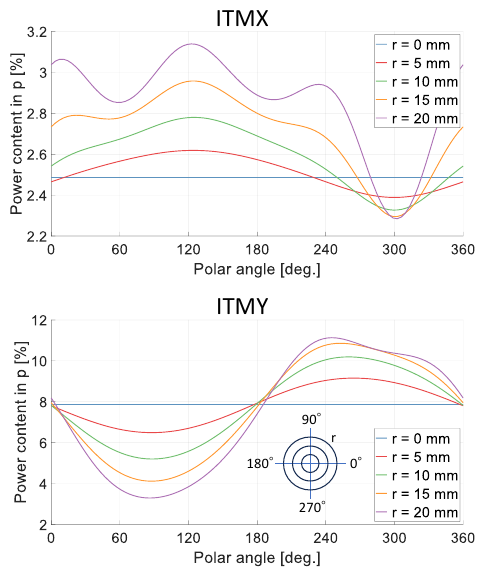}
\caption{Power of p-polarisation when the beam spot is moved away from the mirror centre. The x-axis represents the direction of the decentre and r is the distance between the beam spot centre and mirror centre. Birefringence effect gets smaller when moving the beam towards southeast on ITMX and towards north on ITMY, as shown in Figure~\ref{fig: beamshape}.}
\label{fig: beamposition}
\end{center}
\end{figure}

Another factor that can affect the estimated power is the position of the beam spot on the mirror. The beam spots on the ITMs in KAGRA are not exactly in the centre but are a couple of centimetres away where the power recycling gain and arm transmission power are higher, which is probably due to the inhomogeneous birefringence effect. Figure~\ref{fig: beamposition} shows our estimate of the extent to which the p-polarised beam power changes when the beam is miscentered.

\begin{figure}[htbp]
\begin{center}
\includegraphics[width=8.5cm]{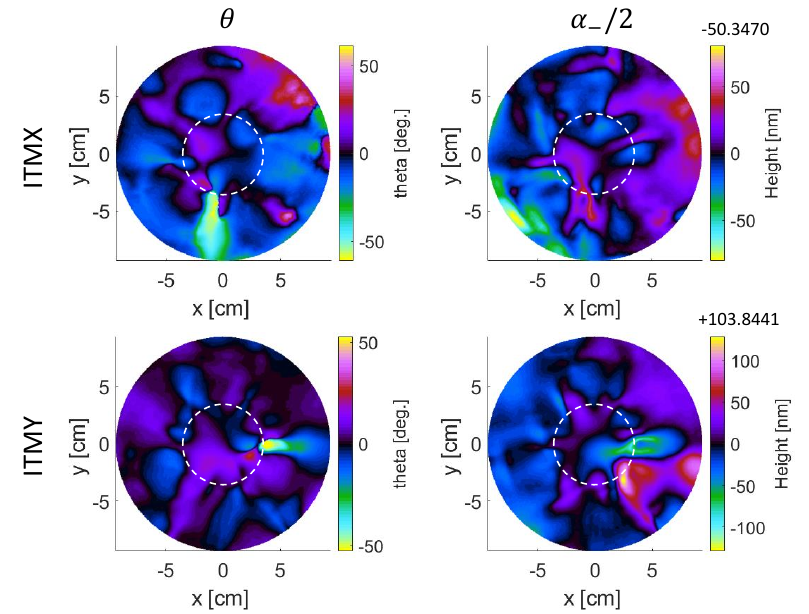}
\caption{Calibrated $\theta$ and $\alpha_-$ maps for ITMX and ITMY. Dashed circles represent the laser beam diameter on mirrors. The plots shown here are the one-way differential phase $\alpha_-/2$. The unit of $\alpha_-$ is nanometre and the laser wavelength 1064~nm corresponds to a radian of $2\pi$.}
\label{fig: finalmap}
\end{center}
\end{figure}

The derived $\theta$ and $\alpha_-$ maps for ITMX and ITMY are shown in Figure~\ref{fig: finalmap}. From these maps, we can see that the birefringence in KAGRA’s current ITMs is inhomogeneous. $\theta$ fluctuates a lot in the range of $\pm 50^{\circ}$ due to the fluctuation of the c-axis. The one-way differential phase (retardation) $\alpha_-/2$ is in the order of 100~nm, which corresponds to $\Delta n = n_\mr{o} - n'_\mr{e} = 6.7 \times 10^{-7}$ (with the thickness of the mirror $d=0.15$~m). If we compare these maps with the beam shape in p-polarisation shown in Figure~\ref{fig: beamshape}, it can be found that the beam shape is mainly determined by the shape of $\theta$ map. This is because $\alpha_-$ is small in our case. The $\sin2\theta$ term plays a more important part than $\sin\alpha_-$ in the amplitude of the field in p-polarisation from Equation~(\ref{eq: V'}).

As discussed above, a good knowledge of the piston and tilt in TWE measurements and the location of the beam spot is necessary to precisely estimate the birefringence effect. For more accurate characterisation, the Fizeau interferometer requires improvements and optimisations for the birefringence study.

\section{Conclusion and discussion}\label{sec: Conclusion}

We established a method for reconstructing the 2D birefringence information of current KAGRA sapphire ITMs from TWE measurements. By combining several TWE measurements with a Fizeau interferometer using linear polarisation at different rotation angles, we successfully constructed a birefringence map over a 200 mm area in diameter. Our estimated beam shape for a single-bounce reflected field in orthogonal polarisation shows a good match with the in situ measurements in KAGRA. 

Note that these TWE measurements were not intended for birefringence characterisations. More accurate reconstruction requires improvements to the previous setup, which should target birefringence studies. One possible improvement could be enabling control of the polarisation rotation angle of the input laser in the Fizeau interferometer rather than rotating the tested mirror, as this will remove the influence of the low-order (piston, tilt, and astigmatism caused by deformation due to gravity) error term in the TWE. For future gravitational wave detectors, birefringence has become an important parameter for their core optics because most potential materials that are suitable for cryogenic operation are birefringent. The birefringence-mapping technique described in this study is fast and simple. This is crucial for future detectors with larger test masses. In addition, the specifications of mirror birefringence require a complete understanding of how birefringence affects the interferometer performance.
The derived birefringence map in this study can be used to study the effects of inhomogeneous birefringence in future detectors and help us gain insights into this knowledge.

\begin{acknowledgments}
We would like to thank Shinji Miyoki and Tomotada Akutsu at KAGRA for their support for the measurements. We also thank Masaki Ando, Kentaro Somiya, Kentaro Komori and Satoru Takano for insightful discussions. This work was supported by JSPS KAKENHI Grant No. JP20H05854, No. 22H01228, No. 23H01205, No. 23K19053, No. 24K00649, and by JST PRESTO Grant No. JPMJPR200B. KAGRA is supported by MEXT, JSPS Leading-edge Research Infrastructure Program, JSPS Grant-in-Aid for Specially Promoted Research 26000005, JSPS Grant-inAid for Scientific Research on Innovative Areas 2905: JP17H06358, JP17H06361 and JP17H06364, JSPS Core-to-Core Program A. Advanced Research Networks, JSPS Grantin-Aid for Scientific Research (S) 17H06133 and 20H05639 , JSPS Grant-in-Aid for Transformative Research Areas (A) 20A203: JP20H05854, the joint research program of the Institute for Cosmic Ray Research, University of Tokyo, National Research Foundation (NRF), Computing Infrastructure Project of Global Science experimental Data hub Center (GSDC) at KISTI, Korea Astronomy and Space Science Institute (KASI), and Ministry of Science and ICT (MSIT) in Korea, Academia Sinica (AS), AS Grid Center (ASGC) and the National Science and Technology Council (NSTC) in Taiwan under grants including the Rising Star Program and Science Vanguard Research Program, Advanced Technology Center (ATC) of NAOJ, and Mechanical Engineering Center of KEK.
\end{acknowledgments}

\appendix

\section{Effective Jones matrix of a non-uniform medium along the optical axis}\label{app: Effective Jones matrix of a non-uniform medium along the optical axis}

In Section~\ref{sec: Generating a birefringence map with a Fizeau interferometer}, we assumed that the transmission of light through a substrate can be described by a single Jones matrix (Equation~(\ref{eq: Jones matrix})). This seems to imply that we also assume the birefringence properties ($\theta$, $n_e$ and $n_o$) to be constant along the depth of the substrate. However, this assumption does not hold for most real substrates. For the sapphire substrates of KAGRA mirrors, for example, the fluctuation of birefringence is supposed to originate from the local misalignment of crystal axes, possibly caused by residual internal stress and lattice defects. Therefore, the birefringence properties must vary not only laterally but also along the depth of the substrate. Thus, the question arises: can we use a single Jones matrix to describe a complex birefringent medium? This section addresses  this question.

\begin{figure}[htbp]
\begin{center}
\includegraphics[width=5.5cm]{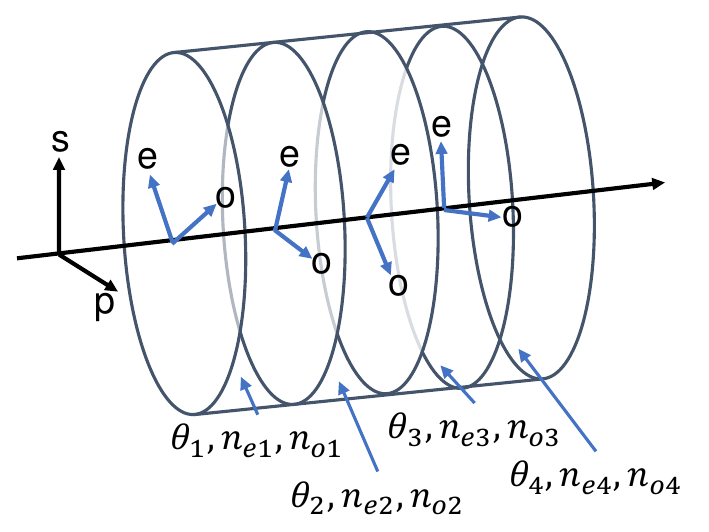}
\caption{Splitting a substrate into thin slices.}
\label{fig: effective Jones matrix}
\end{center}
\end{figure}

We began by splitting a thick substrate into thin slices, as shown in Figure~\ref{fig: effective Jones matrix}. Each slice is sufficiently thin such that the birefringence properties can be regarded as constant. The transmission of light through such a thin slice can be described by a single Jones matrix, as shown in Equation~(\ref{eq: Jones matrix}). 

As we make the slice infinitesimally thin, $\alpha_-$ becomes also infinitesimally small. From Equation~(\ref{eq: Jones matrix}), we can expand the Jones matrix with $\alpha_-$ and leave only its first order terms (the common phase term $\alpha_+$ is omitted here),
\begin{align}
    M &= 
    \begin{pmatrix}
        \cos\alpha_- + i\cos2\theta\sin\alpha_- & -i\sin2\theta\sin\alpha_-  \\
        -i\sin2\theta\sin\alpha_-  & \cos\alpha_- - i\cos2\theta\sin\alpha_-
    \end{pmatrix}\\
    &\sim
    \begin{pmatrix}
            1 + i\cos2\theta\cdot\alpha_- & -i\sin2\theta\cdot \alpha_-  \\
            -i\sin2\theta\cdot \alpha_-  & 1 - i\cos2\theta\cdot \alpha_-
    \end{pmatrix}. \label{eq: approx Jones matrix}   
\end{align}

Next, we considered the transmission of light through two consecutive slices with different Jones matrices: 
\begin{align}
    M_1 &= 
    \begin{pmatrix}
        1 + i\cos2\theta_1\cdot\alpha_1 & -i\sin2\theta_1\cdot \alpha_1  \\
        -i\sin2\theta_1\cdot \alpha_1  & 1 - i\cos2\theta_1\cdot \alpha_1
    \end{pmatrix},\\
    M_2 &= 
    \begin{pmatrix}
        1 + i\cos2\theta_2\cdot\alpha_2 & -i\sin2\theta_2\cdot \alpha_2  \\
        -i\sin2\theta_2\cdot \alpha_2  & 1 - i\cos2\theta_2\cdot \alpha_2
    \end{pmatrix}.   
\end{align}
Such a process can be described by the multiplication of the two Jones matrices,
\begin{widetext}
\begin{align}
    &M_{12} = M_2\cdot M_1 =
    \begin{pmatrix}
        1 + i\left(\cos2\theta_1\cdot\alpha_1 + \cos2\theta_2\cdot\alpha_2\right) & -i\left(\sin2\theta_1\cdot \alpha_1 + \sin2\theta_2\cdot \alpha_2\right)\\
        -i\left(\sin2\theta_1\cdot \alpha_1 + \sin2\theta_2\cdot \alpha_2\right) & 1 - i\left(\cos2\theta_1\cdot\alpha_1 + \cos2\theta_2\cdot\alpha_2\right)
    \end{pmatrix}.\label{eq: combined Jones matrix}
\end{align}
\end{widetext}
By comparing Equation~(\ref{eq: combined Jones matrix}) and Equation~(\ref{eq: approx Jones matrix}), we find $\alpha_\mr{e}$ and $\theta_\mr{e}$, which satisfy the following conditions:
\begin{align}
    \alpha_\mr{e}\cos2\theta_\mr{e} &= \cos2\theta_1\cdot\alpha_1 + \cos2\theta_2\cdot\alpha_2\\
    \alpha_\mr{e}\sin2\theta_\mr{e} &= \sin2\theta_1\cdot\alpha_1 + \sin2\theta_2\cdot\alpha_2.
\end{align}
Finding such a combination of $\alpha_\mr{e}$ and $\theta_\mr{e}$ is always possible. Then we express $M_{12}$ using these parameters as follows:
\begin{equation}
    M_{12} = 
    \begin{pmatrix}
        1 + i\cos2\theta_\mr{e}\cdot\alpha_\mr{e} & -i\sin2\theta_\mr{e}\cdot \alpha_\mr{e}  \\
        -i\sin2\theta_\mr{e}\cdot \alpha_\mr{e}  & 1 - i\cos2\theta_\mr{e}\cdot \alpha_\mr{e}
    \end{pmatrix}.
\end{equation}
This is an effective Jones matrix for transmitting two consecutive thin slices. By repeating this process, we can obtain a single Jones matrix of the form shown in Equation~(\ref{eq: Jones matrix}), which represents the transmission of a thick substrate. Therefore, the use of Equation~(\ref{eq: Jones matrix}) for the analysis of thick, inhomogeneous substrates is justified.

\section{Influence of tilt of TWE maps on the estimation of beam shape and beam power}\label{app: tilt}

In Section~\ref{sec: Constructing birefringence maps from TWE}, we show that the piston, which is typically ignored in interferometric measurements using Fizeau interferometers, is an important factor in birefringence measurements using the method described in this paper. Tilt, which is also ignored in TWE measurements, has a similar effect. In this section, we provide an example of its influence on birefringence characterisation.

\begin{figure}[htbp]
	\begin{center}
		\includegraphics[width=8.5cm]{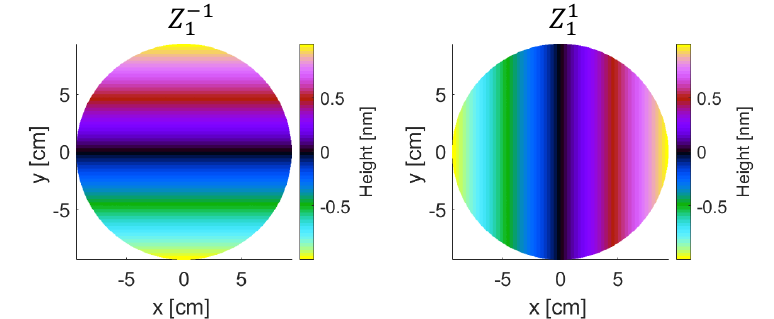}
		\caption{The first order coefficients of Zernike polynomials. Tilts of a surface can be described using a combination of them with different amplitudes.}
		\label{fig: Z1}
	\end{center}
\end{figure}

The tilt of the surface can be described by a combination of the first-order Zernike polynomials $Z_1^{-1}$ and $Z_1^{1}$ (shwon in Figure~\ref{fig: Z1}). Therefore, we need to add two degrees of freedom of tilt for the numerator and denominator in Equation~(\ref{eq: new r}), which results in four more degrees of freedom ($a$, $b$, $c$ and $d$) for the parameter scan, yielding
\begin{equation}\label{eq: r with tilt}
	r = \frac{\psi(\eta_1) - \psi(\eta_2)+p_1+aZ_1^{-1}+bZ_1^{1}}{\psi(\eta_3) - \psi(\eta_4)+p_2+cZ_1^{-1}+dZ_1^{1}}.
\end{equation}

\begin{figure}[htbp]
	\begin{center}
		\includegraphics[width=8.5cm]{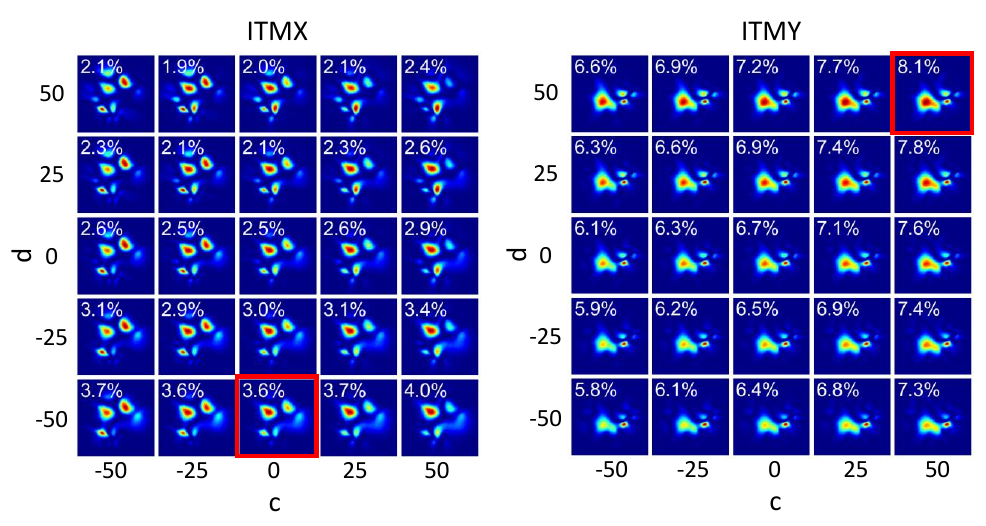}
		\caption{An example of parameter scan showing the change of beam shape and power in p-polarisation with different magnitude of tilt.}
		\label{fig: tilt}
	\end{center}
\end{figure}

According to Figure~\ref{fig: piston}, the change in the numerator of Equation~(\ref{eq: r with tilt}) mainly influences the continuity factor, whereas the change in the denominator has a larger effect on the beam shape and power. Here, we set $a=b=0$ to determine the influence of the tilt on the beam shape and power. Pistons $p_1$ and $p_2$ are fixed according to their values, resulting in the red squares shown in Figure~\ref{fig: piston}. Figure~\ref{fig: tilt} shows the change in the beam shape and power when different amounts of tilt are applied to the denominator. The results labelled with red squares show a good match with the birefringence measurements at the POP in terms of both the beam shape and beam power in p-polarisation.

\bibliography{main}

\end{document}